\documentclass[conference]{IEEEtran}
\IEEEoverridecommandlockouts
\usepackage{cite, balance}
\usepackage{amsmath,amssymb,amsfonts}
\usepackage{enumitem}
\usepackage{graphicx}
\usepackage{textcomp}
\usepackage{subcaption}
\usepackage[utf8]{inputenc}
\usepackage{url}
\usepackage{xcolor}
\usepackage[linesnumbered,ruled,vlined]{algorithm2e}
\def\BibTeX{{\rm B\kern-.05em{\sc i\kern-.025em b}\kern-.08em
    T\kern-.1667em\lower.7ex\hbox{E}\kern-.125emX}}

\begin{document}

\title{Routing Heterogeneous Traffic \\in Delay Tolerant Satellite Networks}

\author{
\IEEEauthorblockN{
Pablo G. Madoery\IEEEauthorrefmark{1}
Gunes Karabulut Kurt\IEEEauthorrefmark{1}\IEEEauthorrefmark{4}
Halim Yanikomeroglu\IEEEauthorrefmark{1}}
Peng Hu\IEEEauthorrefmark{2}
Khaled Ahmed\IEEEauthorrefmark{3}
Guillaume Lamontagne\IEEEauthorrefmark{3}\\

\vspace*{0.3cm}

\IEEEauthorblockA{\IEEEauthorrefmark{1} Department of Systems and Computer Engineering, Carleton University, Ottawa, ON K1S 5B6, Canada}
\IEEEauthorblockA{\IEEEauthorrefmark{2} National Research Council Canada}
\IEEEauthorblockA{\IEEEauthorrefmark{3} Satellite Systems, MDA, Sainte-Anne-de-Bellevue, QC H9X 3R2, Canada}
\IEEEauthorblockA{\IEEEauthorrefmark{4} Poly-Grames Research Center, Department of Electrical Engineering, Polytechnique Montréal, Montréal, QC, Canada}
}

\maketitle
\IEEEpubidadjcol

\begin{abstract}
Delay Tolerant Networking (DTN) has been proposed as a new architecture to provide efficient store-carry-and-forward data transport in satellite networks.
Since these networks relay on scheduled contact plans, the Contact Graph Routing (CGR) algorithm can be used to optimize routing and data delivery performance. 
However, in spite of the various improvements that have been made to CGR, there have been no significant proposals to prioritize traffic with different quality of service requirements. In this work we propose adaptations to CGR that allow performance improvements when sending traffic with different latency constraints, and develop a linear programming optimization model that works as a performance upper bound. The simulation results of the proposed schemes are promising and open the debate on other ways to improve performance while meeting the particular needs of heterogeneous traffic.

\end{abstract}


\section{Introduction}
Satellite networks are becoming increasingly popular as a means to provide high quality imagery, video and communication services around the globe~\cite{7500896}. Efficient space-terrestrial communication technologies, capable of successfully moving large volumes of data between space and ground, are a key element in these networks. In this context, Delay Tolerant Networking (DTN) has been identified as a novel approach that can meet this goal in a cost-effective way by relaxing communication requirements and network infrastructure usually assumed in traditional protocols. The DTN architecture, originated from deep-space and interplanetary networking, embraces the concept of occasionally-connected networks that may suffer from frequent partitions, high delay, and that may be comprised of more than one divergent set of protocols~\cite{RFC4838}.
To this end, a \textit{bundle layer} that exists at a layer above the transport layers of the network, employs a persistent storage on each DTN node to store-carry-and-forward data packets as transmission opportunities become available. 

In the case of satellite networks, the forthcoming episodes of communications and their properties can be determined in advance based on orbital dynamics. These types of deterministic DTNs are known as scheduled DTNs, and can take advantage of a \textit{contact plan} comprising the future network connectivity in order to optimize routing and data forwarding. The Contact Graph Routing (CGR) algorithm, described in \cite{Fraire2021Routing}, is the most developed solution for these networks, and has been the subject of numerous contributions from the research community. These contributions include source routing extensions \cite{edward2011improving}, the adaptation of Dijkstra \cite{segui2011enhancing}, the prevention of routing loops and consideration of multiple destinations \cite{birrane2012analysis, caini2021schedule}, the application of overbooking management techniques \cite{bezirgiannidis2014contact, bezirgiannidis2016analysis}, and congestion mitigation techniques \cite{madoery2018congestion}, route table management strategies and the incorporation of Yen’s algorithm \cite{fraire2018route}. Also, the development of adaptations to improve the scalability \cite{madoery2018managing}, the
incorporation of opportunistic \cite{burleigh2016toward} and probabilistic contacts \cite{RAVERTA2021102663}, a spanning-tree formulation
to compute routes to several destinations \cite{de2019efficient}, and a partial queue information sharing \cite{dhara2019cgr}, among many other contributions.


However, despite the great interest of the community, there have been no significant proposals to adapt CGR in order to prioritize the delivery of traffic that may have heterogeneous requirements in terms of quality of service (QoS). Therefore, in this work our contribution is twofold: on the one hand we propose adaptations to the way CGR chooses the best routes in order to prioritize the traffic according to the required qualities of service (particularly maximum latency), and on the other hand we model the problem with a linear programming optimization model that provides us with a performance bound. 


The rest of the paper is organized as follows.
In Section~\ref{sec:routing schemes} we describe the processes involved when routing in scheduled satellite networks, how CGR carries out the different tasks, and adaptations to CGR that take into account QoS requirements.
In addition, we provide an optimal traffic flow model, and an execution example comparing the behaviour of these 3 schemes in a particular scenario.
Then, in Section~\ref{sec:evaluation} we describe the simulation environment and the results obtained with different performance metrics.
Finally, in Section \ref{sec:futurework} we describe future lines of research and in Section \ref{sec:conclusion} we conclude the work by summarizing the main findings.

\section{Routing Schemes}
\label{sec:routing schemes}

\subsection{Routing in scheduled satellite networks}
\label{sec:routing scheduled}
In delay tolerant satellite networks with planned communications, four processes are usually involved in order to send traffic from the sources to the destinations:

\begin{itemize}
    \item \textbf{Planning}: contact plans are determined by a central entity (a ground station or a Mission Operation Center (MOC)) based on the estimation of future episodes of communication. This task involves taking into account the physical disposition and orientation of nodes through time, as well as their communication system configuration (antenna, modulation, transmission power, etc.). As a result, orbital propagators and communication models are combined to determine a contact plan which can be further tuned to reduce energy consumption or remove conflicting contacts. Then, this contact plan is distributed to satellites in order to run routing processes in a distributed manner when traffic is generated or received by different satellites in the network.
    \item \textbf{Routing}: each satellite uses the contact plan as an input in order to build a routing table. Unlike stable networks such as the Internet, where each route can be thought of as a sequence of nodes, here each route is constructed as a sequence of temporary contacts. Therefore, these routes have associated times during which they remain valid (they are no longer valid when one of their contacts ends). Other attributes that can be calculated for these routes are the delivery time, number of hops / contacts, remaining capacity, etc.
    \item \textbf{Forwarding}: When traffic is generated or received at a satellite, it consults its routing table, performs a filtering of those routes that remain being valid and chooses the one that optimizes a metric that is usually delivery time. Then, by using the information associated with the chosen route, it proceeds to put the traffic in a queue associated with a given neighbor node. 
    \item \textbf{Transmission}: Finally, when contacts occur between satellites, the enqueued traffic is then transmitted in a particular order.
\end{itemize}

\subsection{Contact Graph Routing (CGR)}
\label{sec:cgr}
As explained in \cite{Fraire2021Routing} and summarized here for completeness, CGR is an algorithm that runs in a distributed fashion on different satellites in order to first build a routing table (routing process) and then select routes to enqueue traffic (forwarding process).

In order to construct the route table, it starts from the contact plan received as input, and builds a structure called contact graph, where the vertices correspond to contacts, while the arcs correspond to data retention events in a satellite. Although not so intuitive, the use of this static structure facilitates the execution of route search algorithms in a topology that is actually time-varying. By using this contact graph structure, the search process itself is an adapted version of Dijkstra that explores the different contacts (nodes in the contact graph) and ends up selecting those whose sequence determines the optimal route in terms of earliest delivery time to destination. 
Here we will refer to this scheme as \textit{CGR-DelTime} for comparison purposes.
Furthermore, since routes have a deadline, and they also have a limited capacity, CGR does not compute only 1 route to a destination, but computes the K best routes using Lawler's modification of Yen's algorithm.

Then, in a second stage, during the forwarding process, CGR start from a packet and the routing table built in the previous step, it filters the routes that are no longer valid or that have consumed their capacity, and selects the best route according to the earliest deliver time metric.
After making some annotations of the consumed capacity of the route, this second step is repeated for each new packet that needs to be forwarded.

\subsection{Contact Graph Routing Adaptation}
\label{sec:cgradaptation}

One thing to note here is that CGR already takes into account the time-to-live ($ttl$) of a packet when filtering the routes, and proceeds to ignore those routes that have a projected delivery time later than what is required. However, by choosing a route that is strictly better in terms of delivery time, it is possible that a route with many contacts (hops) will be chosen and that these contacts will also have to send traffic generated by other nodes. Therefore, this can lead to a congestion situation where the route chosen by CGR cannot be honored by downstream nodes, and where the traffic will not be able to reach the destination in the required time. 

In order to try to avoid this situation, we will consider in this work an adaptation to CGR in the forwarding stage.
In particular, this new scheme called \textit{CGR-Hops} will continue filtering those routes that fail to deliver a packet in the required time, as the original \textit{CGR-DelTime} scheme, but it will then choose as the best route the one in which the number of hops is the smallest. The expectation is that this scheme will make decisions that relieve congestion by using fewer hops, and that this congestion relief will serve to leave capacity to the traffic that needs to arrive earlier.

\subsection{Optimal Traffic Flow Model}
\label{sec:lpmodel}

In order to compare different routing strategies, it is useful to have an optimization model that, although it cannot be applied in realistic networks due to computational complexity, serves as a performance upper bound to compare the proposed schemes in smaller networks.

The problem of optimally routing multiple traffics is known in the literature as \textit{multi-commodity flow} \cite{Hu:1963}. In the context of DTN networks, the works \cite{Jain:2004, Alonso03,fraire2016-tacp} presented linear programming models that allow solving this problem in networks with intermittent and scheduled communications. Here, we will describe a new version of these models tailored to our particular needs and extended to consider traffics with different quality of service in terms of latency.

\subsubsection{Coefficients and Decision Variables}

The topology is discretized into $K$ states, where each state $k_q \in K$ represents the network during a time interval $[t_{q-1}, t_{q}]$.
Thus, the network is defined by $k_1,...,k_q,...k_f$ states,
which correspond to time intervals $[t_{0},t_{1}],...,[t_{q-1},t_{q}],...,[t_{f-1},t_{f}]$.
Within each state $k_q$, the network is represented by a graph $G_{k_q}$ whose vertices $v \in V$ correspond to nodes and whose arcs $e \in E_{k_q}$ correspond to communication opportunities (contacts) between nodes in $k_q$. 
The notation $c_{e}$ represents the capacity in number of packets that can be sent from the source node to the destination node of an arc $e$ during one state $k_q$. 
In addition, we represent the number of packets that can be stored in node $v$ buffer as $b_v$.  
In order to use a compact notation we will call $I^v$ the set of arcs $e$ entering a node $v$, while we will call $O^v$ the set of arcs leaving it. Additionally $D$ constitutes the set of all the traffic demands $d_{t_q, ttl}^{y,z}$ generated at time $t_q$ from node $y$ to node $z$ with time-to-live $ttl$. We assume in general that all traffic demands are generated at the beginning of some state. If this is not the case, it is always possible to add a new intermediate state to contemplate the generation of such traffic. On the other hand, within the output variables, $X_{e}^{y,z}$ constitutes the number of packets of traffic $y,z$ that are sent on arc $e$, while the variable $B_{v, t_q}^{y,z}$ is the number of packets of traffic $y,z$ stored at node $v$ at time $t_q$. We summarize these model parameters in Table \ref{Tab-LP}.

\begin{table}[t]
\centering
\caption{Optimal Traffic Flow Model Parameters}
\label{Tab-LP}
\begin{tabular}{|c|c|}
\multicolumn{2}{c}{Input Coefficients}                                                                                                 \\ 
\hline
$t_q \in T$              & Timestamps                                                                                                  \\ 
\hline
$k_q =[t_q-1,t_q] \in K$ & States (Time intervals)                                                                                     \\ 
\hline
$v \in V$                & Nodes                                                                                                       \\ 
\hline
$e \in E_{k_q}$            & Arcs of a graph in state $k_q$                                                                              \\ 
\hline
$c_e$                    & Capacity of arc $e$                                                                                         \\ 
\hline
$b_v$                    & Capacity of buffer's node $v$                                                                               \\ 
\hline
$d_{t_q, ttl}^{y,z} \in D$        & \begin{tabular}[c]{@{}c@{}}Traffic from node $y$ to node $z$ \\originated at timestamp $t_q$ \\with time-to-live $ttl$\end{tabular}   \\ 
\hline
\multicolumn{1}{l}{}     & \multicolumn{1}{l}{}                                                                                        \\
\multicolumn{2}{c}{Output Variables}                                                                                                   \\ 
\hline
$\{X_e^{y,z}\}$            & \begin{tabular}[c]{@{}c@{}}Traffic from node $y$ to node $z$ \\sent in arc $e$\end{tabular}                 \\ 
\hline
$\{B_{t_q,v}^{y,z}\}$        & \begin{tabular}[c]{@{}c@{}}Buffer occupancy of node $v$ \\at timestamp $t_q$ by traffic $y,z$\end{tabular}  \\
\hline
\end{tabular}
\end{table}

\subsubsection{Objective Function and Constraints}

Taking into account the coefficients and decision variables defined, we link them by means of the objective function defined in \eqref{LPModel01}
subject to the constraints \eqref{LPModel02} to \eqref{LPModel07}.

\begin{flalign}
	\label{LPModel01}
	\textrm{minimize:} \quad \sum\limits_{k_q \in K} \sum\limits_{e \in E_{k_q}} \sum\limits_{y \in V} \sum\limits_{z \in V} w(k_q) * X_{e}^{y,z} 
\end{flalign}
Subject to:
\begin{equation}
\label{LPModel02}
B_{t_{q},v}^{y,z} = \begin{cases}
B_{t_{q-1},v}^{y,z} + \sum\limits_{e \in I^v} X_{e}^{y,z} - \sum\limits_{e \in O^v} X_{e}^{y,z}  + d_{t_q,ttl}^{y,z} & \text{if $y = v$}\\
B_{t_{q-1},v}^{y,z} + \sum\limits_{e \in I^v} X_{e}^{y,z} - \sum\limits_{e \in O^v} X_{e}^{y,z}  & \text{if $y \neq v$}
\end{cases}
\end{equation}
\begin{flalign}
\label{LPModel03}
\sum\limits_{y \in V} \sum\limits_{z \in V} B_{t_q,v}^{y,z} &<= b_{v} \quad \forall \; t_q,v\\
\label{LPModel04}
\sum\limits_{y \in V} \sum\limits_{z \in V} X_{e}^{y,z} &<= c_{e} \quad \forall \; k_q, e
\end{flalign}
\begin{equation}
\label{LPModel05}
B_{t_0,v}^{y,z} = \begin{cases}
d_{t0,ttl}^{y,z} & \text{if $y = v$}\\
\quad 0 & \text{if $y \neq v$}
\end{cases}
\quad \forall \; v,y,z
\end{equation}
\begin{equation}
\label{LPModel06}
B_{t_q,v}^{y,z} >= \begin{cases}
d_{t_q,ttl}^{y,z} & \text{if $y = v$}\\
\quad 0 & \text{if $y \neq v$}
\end{cases}
\quad \forall \; t_0<t_q, v,y,z
\end{equation}
\begin{equation}
\label{LPModel07}
B_{t_f,v}^{y,z} = \begin{cases}
\sum\limits_{t_q \in T} d_{t_q, ttl}^{y,z} & \text{if $z = v$}\\
\quad 0 & \text{if $z \neq v$}
\end{cases}
\quad \forall \; v,y,z
\end{equation}

A linear programming model is thus formed whose input consists of a set of traffics ($d_{t_q}^{y,z}$) to be sent from sources ($y$) to destinations ($z$), with maximum latency requirements ($ttl$), through a time-varying topology, with buffer capacities ($b_v$) and transmission  capacities ($c_e$), and whose output consists of the optimal flows ($X_e^{y,z}$) to be followed by the different traffics.

The objective function defined in \eqref{LPModel01}
seeks to minimize the sum of the products $w(k_q) * X_{e}^{y,z}$, where $w(k_q)$ is a weighting function that assigns an increasing weight to each state.
Thus, the higher the increment caused by $w$, the higher the cost of using later arcs in time, i.e., minimizing the product prioritizes delivering traffic as soon as possible. On the other hand, the lower the increasing cost of $w$, the more importance is given to using as few arcs as possible.

With respect to the constraints, equation \eqref{LPModel02} defines the buffer occupancy of each node ($v$) at each timestamp ($t_q$) taking into account the state at the previous timestamp ($t_{q-1}$), the incoming ($e \in I^v$) and outgoing ($e \in O^v$) flows in the interval $[t_{q-1}, t_q]$, and the traffic generated at $t_q$ ($d_{t_q, ttl}^{y,z}$).
In turn, equation \eqref{LPModel03} limits the buffer occupancy ($B_{t_q,v}^{y,z}$) of each node according to $b_v$, while equation \eqref{LPModel04} limits the amount of traffic that can flow through each arc ($X_{e}^{y,z}$) according to $c_e$.

In addition, the constraint \eqref{LPModel05}
establishes the buffers at the initial timestamp ($t_0$), while the constraint \eqref{LPModel06} ensures that traffics generated in intermediate states are not sent in earlier states. 
This constraint also ensures that traffics that have a maximum latency constraint ($ttl$) arrive at their destinations on time and remain there until the final state.
Finally, equation \eqref{LPModel07} ensures that all $y,z$ traffics generated over time ultimately reside in their corresponding buffers at the destination nodes ($B_{t_f,v}^{y,z}$).

\subsection{Execution Example}
\label{sec:example}

\begin{figure*}[t!]
\centering
\begin{subfigure}{0.31\textwidth}
    \centering
    \fbox{\includegraphics[width=\textwidth]{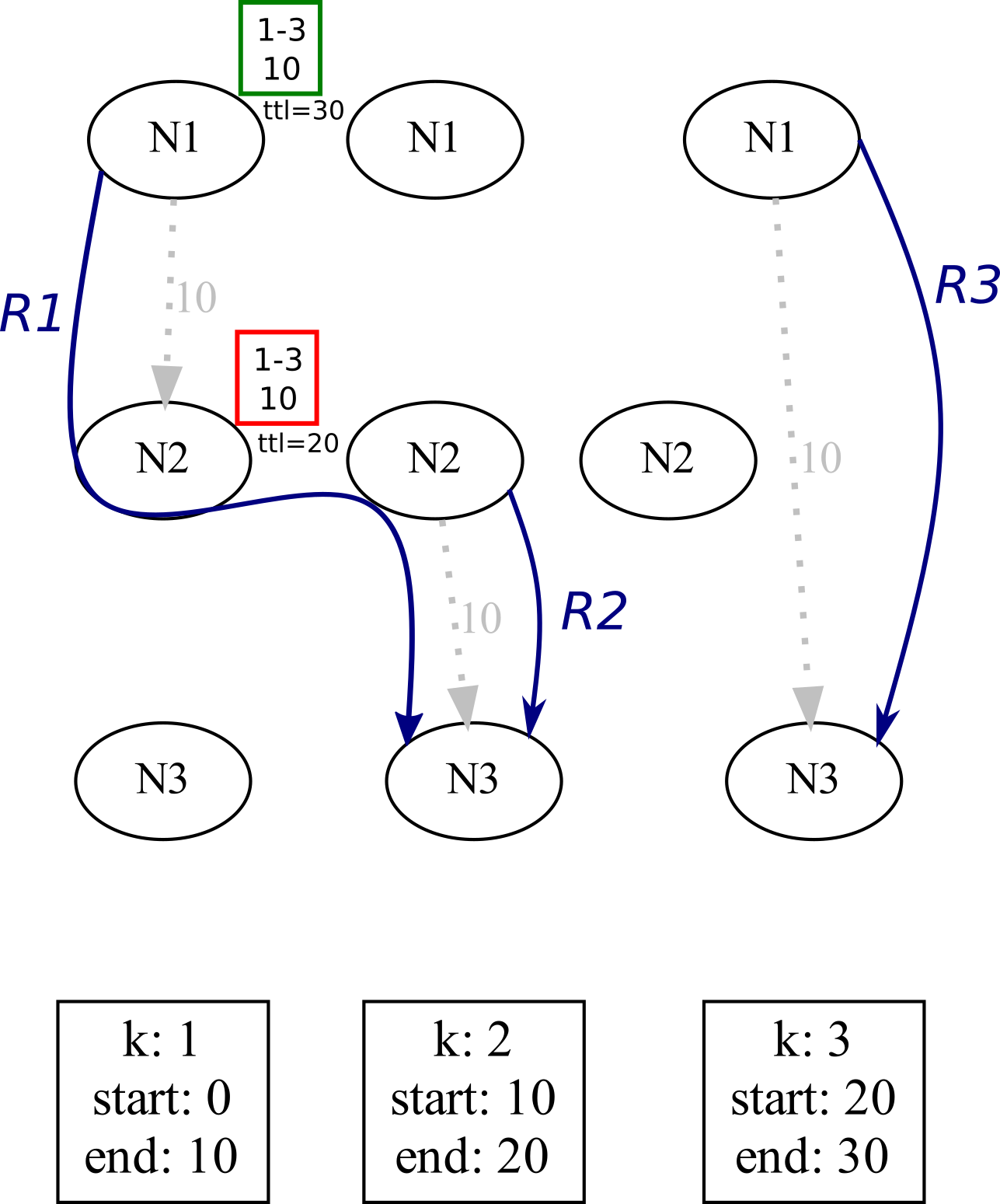}}
    \caption{Topology, traffic and routes}
    \label{fig:Topology}
\end{subfigure}
\hfill
\begin{subfigure}{0.274\textwidth}
    \centering
    \fbox{\includegraphics[width=\textwidth]{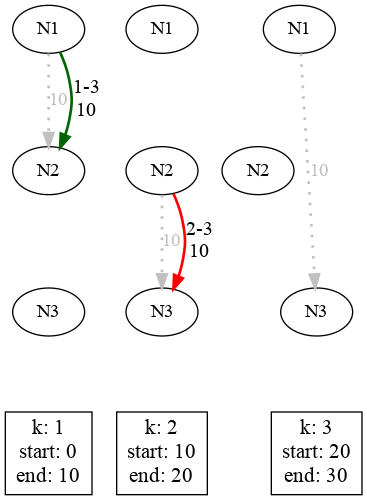}}
    \caption{\textit{CGR-DelTime} traffic flow}
    \label{fig:CGR-DelTime}
\end{subfigure}
\hfill
\begin{subfigure}{0.318\textwidth}
    \centering
    \fbox{\includegraphics[width=\textwidth]{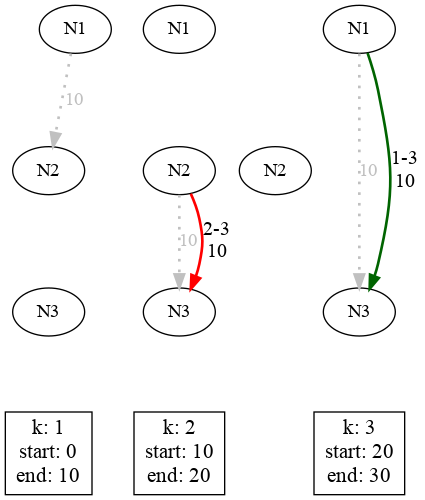}}
    \caption{\textit{LP Model} and \textit{CGR-Hops} traffic flow}
    \label{fig:Model}
\end{subfigure}
\caption{Example of execution of different routing schemes}
\label{fig:example}
\end{figure*}

In order to intuitively understand the behavior of the different routing schemes, we propose to analyze a simple example where 3 nodes ($N1$, $N2$, $N3$) have 3 contacts in different states as shown in Figure \ref{fig:Topology}. 

The dotted lines represent contacts with the capacity to send 10 packets each.
Both nodes $N1$ and $N2$ generate 10 packets in state $k1$ destined for $N3$ (enclosed in squares).
The traffic generated in $N1$ has a $ttl$ of 30 seconds, so it must reach its destination in state $k3$ or earlier, while the traffic generated in $N2$ has a $ttl$ of 20 seconds, so it must reach destination in state $k2$ or earlier. Regarding routes (represented as solid blue lines), $N1$ is able to find routes $R1$ and $R3$ to reach node $N3$, while $N2$ is able to find the route $R2$.

In Figure \ref{fig:CGR-DelTime}, we observe the behavior of \textit{CGR-DelTime} where traffic transmissions are represented with colored solid lines.
Here, N1 calculates and sends the traffic trying to use the route $R1$, however, when the 10 packets arrive at $N2$, they find an unforeseen situation: $N2$ also has traffic for $N3$ and needs to use the full capacity of the $N2-N3$ contact. 
This generates a situation of congestion that causes in turn that the 10 packets of $N1$ cannot reach $N3$ in the required time. On the other hand, the $N2$ traffic does reach $N3$ on time by using the route $R2$.

Finally, Figure \ref{fig:Model} shows the behavior obtained both with  \textit{CGR-Hops} and with the \textit{LP Model}. Here, although \textit{CGR-Hops} is aware that route $R1$ delivers the packets earlier, it chooses route $R3$ which, in addition to delivering the packets on time ($ttl=30$), uses fewer hops.
This congestion avoidance effect obtained by both the \textit{LP Model} and \textit{CGR-Hops} leads to more QoS compliant traffic flows and makes it possible for all traffic to reach its destination on time. The difference between the \textit{LP Model} and \textit{CGR-Hops} in this case is that the \textit{LP Model} needed and used global knowledge of the topology and the traffics to be generated, while \textit{CGR-Hops} was able to have a similar behavior without knowing the traffic generated in other nodes of the network. 

In order to assess whether this approach is promising, we proceed to evaluate and analyze whether this type of behavior continues to occur in more complex cases.

\section{Evaluation}
\label{sec:evaluation}

To compare the performance of different routing schemes, we have adapted and extended Dtnsim. DtnSim is a discrete event-driven simulator developed in the Omnet++ framework, presented in~\cite{Fraire:2017:DtnSim} and made available to be used as free software\footnote{Public DtnSim repository:~\url{https://bitbucket.org/lcd-unc-ar/dtnsim}}.

By using Dtnsim, we have generated 25 random network topologies of 100s divided in 10 states of 10s.
Each topology captures the time-varying connectivity among 11 nodes, where the first 10 nodes represent satellites, while node 11 represents a ground station. 
The connectivity between nodes is based on a contact density parameter $\delta$ which can take values between 0.0 and 1.0. 
A network with $\delta$ = 1.0 is fully connected with contacts present in all states, while $\delta$ = 0 implies no contacts exist at all. We have considered $\delta=0.2$ for the cases under study. The effect of increasing $\delta$ is that higher values of traffic generation are needed to observe the same congestion and performance degradation.
Each arc has a capacity to send 10 traffic packets while there is no limit to the storage capacity at the nodes.

In order to generate congestion situations, we have set up an all-to-one traffic pattern in which all 10 nodes generate a variable traffic load destined to node 11.
In particular, nodes 1 to 5 generate packets that do not have a maximum latency requirement ($ttl$), so they can arrive at their destination at any state, while nodes 6 to 10 generate packets that have a maximum latency requirement of 20 seconds, so they must arrive at the destination in the first state (k=1) or the second state (k=2).

The choice of these parameters has been made in such a way that heterogeneous traffic must be routed, that congestion is eventually provoked, and that the packets have a chance to reach their destination before the end of the simulation.  

\begin{figure*}[t!]
	\centering
	\includegraphics[width=0.9\textwidth]{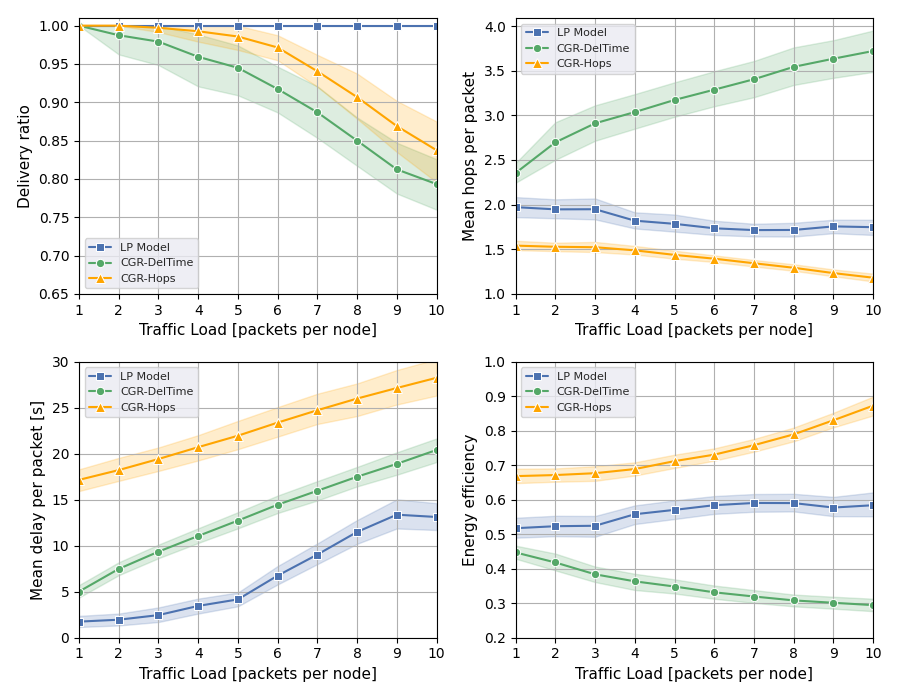}
	\caption{Simulation results}
	\label{results}
\end{figure*}

We compare the performance of \textit{CGR-DelTime}, \textit{CGR-Hops}, and the \textit{LP Model} described in section \ref{sec:routing schemes}. In particular, we analyze the following metrics:
\begin{itemize}
    \item \textit{\textbf{Delivery ratio}}: total number of packets generated divided by the total number of packets arriving at destination that meet their specified maximum latency.
    \item \textit{\textbf{Mean hops per packet}}: total number of transmissions of all packets divided by the total number of packets arriving at destination that meet their specified maximum latency. 
    \item \textit{\textbf{Mean delay per packet}}: average latency with which packets are delivered to the destination meeting their specified maximum latency.
    \item \textit{\textbf{Energy efficiency}}: total number of packets arriving at destination that meet their specified maximum latency divided by the total number of transmissions of all packets.
\end{itemize}

As it can be seen in Figure \ref{results}, regardless of the traffic load, the \textit{LP Model} is able to deliver all generated packets to the destination node while meeting the maximum latency requirement imposed on each packet.  On the other hand, \textit{CGR-DelTime} shows a delivery ratio that decreases as the traffic load increases. This is because the congestion generated starts to prevent the delivery of traffic to the destination. It should be highlighted that when a packet cannot be sent to its destination on time, CGR deletes that packet. Finally, although \textit{CGR-Hops} also experiences a decrease in delivery ratio as the traffic load increases, this decrease is smaller due to the better congestion avoidance capability of this scheme.

Furthermore, \textit{CGR-Hops} shows a favorable performance in terms of the average number of hops per packet because it chooses routes with fewer hops. In this case, the LP Model uses a higher number of hops per packet because, in addition to being constrained to comply with maximum latencies, it seeks to deliver packets as soon as possible regardless of whether it must use more hops to do so. The decreasing trend as the traffic load increases is explained by the fact that when congestion increases, the model starts to give up sending traffic in the first states and use later states with shorter routes. 

Energy efficiency can be interpreted in our case as how many packets are successfully delivered to destination with each transmission. It is the inverse of the mean hops per packet metric although it has the advantage of being limited in range between 0 and 1. It can be seen here that \textit{CGR-Hops} is the most efficient scheme and that the \textit{LP Model} lies between \textit{CGR-DelTime} and \textit{CGR-Hops}. The explanation as to why the model is not the most efficient here lies in the fact that the model seeks to deliver traffic as quickly as possible and ends up using many arcs in the early states before moving on to using arcs in later states. In contrast, \textit{CGR-Hops} proceeds to use routes with fewer hops as long as those routes meet the maximum latency requirement.

Finally, the \textit{LP Model} obtains the best metric in terms of delay per packet followed by \textit{CGR-DelTime} and then by \textit{CGR-Hops}. It is worth noting 2 points here: on the one hand, this metric is encompassing all generated packets regardless of their maximum required latency.  If the average delay is calculated separately for each type of traffic we can expect the average delay per packet to be lower for traffics with lower $ttl$ values.
On the other hand, this metric is calculated only on the delivered packets that meet the maximum required latency, so it should be noted that these schemes are not directly comparable if the delivery ratio is different as in this case. 

In general, the effect that \textit{CGR-Hops} ends up achieving is to delay traffic that can be delayed, thus relieving congestion, and allowing traffic that need to be delivered earlier to reach its destination on time.


\section{Future work}
\label{sec:futurework}
As future work we envision further modifications to CGR to achieve even better performance. In particular:
\begin{itemize}
\item \textit{load balancing}: if all nodes always use the shortest routes, it is likely that
the different routes will share more links, generating a bottleneck in terms of capacity (congestion). The proposal then is not to use only the shortest routes, but that a node can perform a load-balancing by
round-robin on the K best routes.
\item \textit{centrality}: in graph theory, the centrality metric is related to  how many routes pass through a link or node. The idea here would be to use routes through links or nodes that do not have the highest centrality, trying to avoid congestion.
\item \textit{machine learning}: the proposal here is to use machine learning models to estimate the congestion that may exist on a link or node at a given time, and then use these estimates to carry out the routing seeking to avoid those routes more congested but without underutilizing them.
\end{itemize}

\section{Conclusion}
\label{sec:conclusion}
In this work we have proposed and evaluated an adaptation to CGR to improve performance when sending traffic with different maximum latency requirements. The results show that the current capability of CGR to honor latency requirements is limited because it uses only local traffic information, and strictly chooses routes that optimize the expected delivery time without considering the number of hops involved. However, these decisions do not take into account that intermediate nodes can also generate traffic producing congestion situations that prevent the effective fulfillment of latency requirements. On the other hand, the proposed adaptation selects, within the routes that meet the maximum latency requirements, those that have fewer hops. This allows to alleviate congestion and increase the number of packets that are delivered on time.

\section*{Acknowledgments}
This work has been supported in part by the National Research Council Canada's (NRC) High Throughput Secure Networks program within the Optical Satellite Communications Consortium Canada (OSC) framework

\balance
\bibliographystyle{IEEEtran}
\bibliography{biblio}

\begin{thebibliography}{10}
\providecommand{\url}[1]{#1}
\csname url@samestyle\endcsname
\providecommand{\newblock}{\relax}
\providecommand{\bibinfo}[2]{#2}
\providecommand{\BIBentrySTDinterwordspacing}{\spaceskip=0pt\relax}
\providecommand{\BIBentryALTinterwordstretchfactor}{4}
\providecommand{\BIBentryALTinterwordspacing}{\spaceskip=\fontdimen2\font plus
\BIBentryALTinterwordstretchfactor\fontdimen3\font minus
  \fontdimen4\font\relax}
\providecommand{\BIBforeignlanguage}[2]{{%
\expandafter\ifx\csname l@#1\endcsname\relax
\typeout{** WARNING: IEEEtran.bst: No hyphenation pattern has been}%
\typeout{** loaded for the language `#1'. Using the pattern for}%
\typeout{** the default language instead.}%
\else
\language=\csname l@#1\endcsname
\fi
#2}}
\providecommand{\BIBdecl}{\relax}
\BIBdecl

\bibitem{7500896}
J.~Alvarez and B.~Walls, ``Constellations, clusters, and communication
  technology: Expanding small satellite access to space,'' in \emph{2016 IEEE
  Aerospace Conference}, March 2016, pp. 1--11.

\bibitem{RFC4838}
V.~Cerf, S.~Burleigh, A.~Hooke, L.~Torgerson, R.~Durst, K.~Scott, K.~Fall, and
  H.~Weiss, ``Delay-tolerant networking architecture,'' Internet Requests for
  Comments, RFC Editor, RFC 4838, April 2007.

\bibitem{Fraire2021Routing}
J.~A. Fraire, O.~{De Jonckère}, and S.~C. Burleigh, ``Routing in the space
  internet: A contact graph routing tutorial,'' \emph{Journal of Network and
  Computer Applications}, vol. 174, p. 102884, 2021.

\bibitem{edward2011improving}
J.~Edward, ``Improving graph-based overlay routing in delay tolerant
  networks,'' \emph{IEEE Wireless Days}, 2011.

\bibitem{segui2011enhancing}
J.~Segui, E.~Jennings, and S.~Burleigh, ``Enhancing contact graph routing for
  delay tolerant space networking,'' in \emph{IEEE Global Telecommunications
  Conference}, 2011, pp. 1--6.

\bibitem{birrane2012analysis}
E.~Birrane, S.~Burleigh, and N.~Kasch, ``Analysis of the contact graph routing
  algorithm: Bounding interplanetary paths,'' \emph{Acta Astronautica},
  vol.~75, pp. 108--119, 2012.

\bibitem{caini2021schedule}
C.~Caini, G.~M. De~Cola, and L.~Persampieri, ``Schedule-aware bundle routing:
  Analysis and enhancements,'' \emph{International Journal of Satellite
  Communications and Networking}, vol.~39, no.~3, pp. 237--249, 2021.

\bibitem{bezirgiannidis2014contact}
N.~Bezirgiannidis, C.~Caini, D.~P. Montenero, M.~Ruggieri, and V.~Tsaoussidis,
  ``Contact graph routing enhancements for delay tolerant space
  communications,'' in \emph{7th IEEE advanced satellite multimedia systems
  conference and the 13th signal processing for space communications workshop
  (ASMS/SPSC)}, 2014, pp. 17--23.

\bibitem{bezirgiannidis2016analysis}
N.~Bezirgiannidis, C.~Caini, and V.~Tsaoussidis, ``Analysis of contact graph
  routing enhancements for dtn space communications,'' \emph{International
  Journal of Satellite Communications and Networking}, vol.~34, no.~5, pp.
  695--709, 2016.

\bibitem{madoery2018congestion}
P.~G. Madoery, J.~A. Fraire, and J.~M. Finochietto, ``Congestion management
  techniques for disruption-tolerant satellite networks,'' \emph{International
  Journal of Satellite Communications and Networking}, vol.~36, no.~2, pp.
  165--178, 2018.

\bibitem{fraire2018route}
J.~A. Fraire, P.~G. Madoery, A.~Charif, and J.~M. Finochietto, ``On route table
  computation strategies in delay-tolerant satellite networks,'' \emph{Ad Hoc
  Networks}, vol.~80, pp. 31--40, 2018.

\bibitem{madoery2018managing}
P.~G. Madoery, J.~A. Fraire, F.~D. Raverta, J.~M. Finochietto, and S.~C.
  Burleigh, ``Managing routing scalability in space {DTNs},'' in \emph{6th IEEE
  International Conference on Wireless for Space and Extreme Environments
  (WiSEE)}, 2018, pp. 177--182.

\bibitem{burleigh2016toward}
S.~Burleigh, C.~Caini, J.~J. Messina, and M.~Rodolfi, ``Toward a unified
  routing framework for delay-tolerant networking,'' in \emph{IEEE
  International Conference on Wireless for Space and Extreme Environments
  (WiSEE)}, 2016, pp. 82--86.

\bibitem{RAVERTA2021102663}
F.~D. Raverta, J.~A. Fraire, P.~G. Madoery, R.~A. Demasi, J.~M. Finochietto,
  and P.~R. D’Argenio, ``Routing in delay-tolerant networks under uncertain
  contact plans,'' \emph{Ad Hoc Networks}, vol. 123, p. 102663, 2021.

\bibitem{de2019efficient}
O.~De~Jonck{\`e}re, ``Efficient contact graph routing algorithms for unicast
  and multicast bundles,'' in \emph{IEEE International Conference on Space
  Mission Challenges for Information Technology (SMC-IT)}, 2019, pp. 87--94.

\bibitem{dhara2019cgr}
S.~Dhara, C.~Goel, R.~Datta, and S.~Ghose, ``{CGR-SPI}: A new enhanced contact
  graph routing for multi-source data communication in deep space network,'' in
  \emph{IEEE International Conference on Space Mission Challenges for
  Information Technology (SMC-IT)}, 2019, pp. 33--40.

\bibitem{Hu:1963}
T.~C. Hu, ``Multi-commodity network flows,'' \emph{Oper. Res.}, vol.~11, no.~3,
  pp. 344--360, Jun. 1963.

\bibitem{Jain:2004}
S.~Jain, K.~Fall, and R.~Patra, ``Routing in a delay tolerant network,'' in
  \emph{Proceedings of the 2004 Conference on Applications, Technologies,
  Architectures, and Protocols for Computer Communications}, ser. SIGCOMM
  '04.\hskip 1em plus 0.5em minus 0.4em\relax New York, NY, USA: ACM, 2004, pp.
  145--158.

\bibitem{Alonso03}
J.~Alonso and K.~Fall, ``A linear programming formulation of flows over time
  with piecewise constant capacity and transit times,'' Tech. Rep., 2003.

\bibitem{fraire2016-tacp}
J.~A. Fraire, P.~G. Madoery, and J.~M. Finochietto, ``Traffic-aware contact
  plan design for disruption-tolerant space sensor networks,'' \emph{Ad Hoc
  Networks}, vol.~47, pp. 41 -- 52, 2016.

\bibitem{Fraire:2017:DtnSim}
J.~A. Fraire, P.~G. Madoery, F.~Raverta, J.~M. Finochietto, and R.~Velazco,
  ``Dtnsim: Bridging the gap between simulation and implementation of
  space-terrestrial dtns,'' in \emph{Space Mission Challenges for Information
  Technology (SMC-IT), 2017 IEEE Int. Conference on}, Sept 2017.

\end{thebibliography}

\end{document}